\documentclass[
twocolumn,
]{revtex4-2}
\usepackage[T1]{fontenc}

\usepackage[
    a4paper, 
    mag=1000,
    left=1.5cm, 
    right=1.5cm,
    top=2cm,
    bottom=2cm,
    headsep=0.7cm,
    footskip=1cm
    ]{geometry}
    
\usepackage[utf8]{inputenc}
\usepackage{amssymb}
\usepackage{mathtools}
\usepackage[T2A]{fontenc}
\usepackage{amsmath}
\usepackage[russian,english]{babel}
\usepackage{graphicx}
\usepackage[table]{xcolor}

\usepackage{amsfonts} 

\usepackage{enumerate}

\usepackage{graphicx}% Include figure files
\usepackage{dcolumn}% Align table columns on decimal point
\usepackage{bm}% bold math
\usepackage{hyperref}% add hypertext capabilities
\usepackage{array}
\usepackage{comment}
\usepackage{float}
\usepackage{subcaption}
\usepackage{xcolor}

\begin{document}

\title{Low-complexity neural network equalization for long-haul coherent transmission with cascaded semiconductor optical amplifiers}

\author{S.~Bogdanov$^{1*}$, S.~Sygletos$^{1}$, O.~Sidelnikov$^{2}$, G.~Gomes$^{1}$, M.~Kamalian-Kopae$^{3}$ and S. K. Turitsyn$^{1}$} 

\affiliation{$^{1}$Aston Institute of Photonic Technologies, Aston University, Birmingham, B4 7ET, UK}
\affiliation{$^{2}$Novosibirsk State University, Novosibirsk, 630090, Russia}
\affiliation{$^{3}$Microsoft Azure Fiber, Unit 7, The Quadrangle, Romsey, SO561 9DL, UK}
\affiliation{$^{*}${s.bogdanov@aston.ac.uk}}

\begin{abstract} 
In this letter, we numerically investigate a long-haul coherent data transmission system with a cascade of semiconductor optical amplifiers (SOAs). We exploit low-complexity neural networks that can be implemented in real time to compensate for the accumulated distortions induced by a cascade of SOAs. This equalization provides an order-of-magnitude reduction in bit error rate at low dispersion (in the O-band), whereas higher dispersion degrades performance.
\end{abstract}

%\setboolean{displaycopyright}{false} % Do not include copyright or licensing information in submission.

\maketitle

\section{Introduction}

Semiconductor optical amplifiers (SOAs) possess several attractive properties for fiber optic communications. They are cost-effective, compact, and integrable with existing data transmission systems, while offering a broad amplification bandwidth (typically $\sim 30$--$80\,\mathrm{nm}$ at the 3-dB level) \cite{connelly2007semi, sobhanan2022semiconductor}. This wide bandwidth is particularly important as ultra-wideband transmission remains a dominant route to higher link capacity, making SOAs a compelling alternative to complex multi-stage amplification schemes. For example, amplification across the full low-loss window of standard fibre (O--E--S--C--L--U bands) has been demonstrated using a cascade of six amplifiers of different types, including Raman amplification \cite{puttnam2024402}. In contrast, semiconductor devices can potentially provide gain across these bands within a single technology platform. This capability is particularly relevant for ultra-wideband transmission fibres, such as hollow-core fibres, which can offer low-loss windows over a wide spectral range.
Moreover, the growing demand for data-centre interconnections requires cost-effective and reliable amplification in the O-band, where conventional erbium-doped fibre amplifiers are ineffective \cite{agrell2024roadmap}. Although O-band transmission currently relies on specialised bismuth-doped fibre amplifiers, SOAs can offer a compact alternative while retaining the advantages listed above \cite{elson2024continuous}. Finally, because SOAs are electrically pumped, they avoid the need for optical pump lasers, simplifying deployment and enabling low-power, highly integrable amplification.

Nonetheless, SOAs exhibit several limitations, including signal distortions arising from gain saturation and the finite carrier lifetime, polarization-dependent gain, and relatively high noise figures \cite{sobhanan2022semiconductor}. The finite gain recovery time, when comparable to or longer than the symbol period, gives rise to pattern effects because the gain does not fully recover from one symbol to the next \cite{mecozzi1997saturation}, leading to symbol-to-symbol power fluctuations. In coherent transmission, the temporally varying gain can also translate into nonlinear phase noise via the linewidth enhancement factor \cite{agrawal1989self}. These effects motivate system designs that are robust to SOA impairments, and a range of mitigation schemes has been proposed.
One approach is to linearize the SOA response by modulating the drive current using the input data stream \cite{saleh1988compensation, ortiz2020numerical}. Alternatively, a simplified linear backpropagation procedure was described in \cite{ghazisaeidi2011efficiency}, with the authors reporting that it outperforms the standard fourth-order Runge–Kutta implementation as the sampling rate approached realistic Nyquist values. More recently, a trainable digital filter was applied to compensate SOA-induced nonlinearities in a $\mathrm{32\, Gbaud}$ 16-QAM/64-QAM back-to-back system \cite{sillekens2023experimental}, achieving performance comparable to conventional methods, but with reduced parameter-estimation time.

In cascaded SOA systems, impairments tend to accumulate, leading to substantial penalties in long-haul, multi-span links. An experimental study of metro and access networks employing a cascade of four SOAs reported progressively increasing degradation along the cascade \cite{koenig2014amplification}. Recently, long-haul experiments using an SOA for fiber-loss compensation demonstrated a lower capacity than an EDFA-only scheme \cite{sillekens2025experimental}. Nevertheless, potential advantages such as reduced energy consumption and broad wavelength applicability motivate SOA-based solutions when combined with appropriate digital signal processing (DSP).

Machine-learning methods, and neural networks (NNs) in particular, have been extensively studied for equalization in fiber-optic channels and have demonstrated strong performance \cite{musumeci2018overview, khan2019optical, freire2022neural}. These approaches have subsequently been applied to SOA-based amplification systems. For instance, a two-cascaded  $100\,\mathrm{Gbit/s}$ PAM-4 IM/DD O-band link was investigated in \cite{wang2022mitigation}. Among multilayer perceptrons (MLPs), convolutional neural networks (CNNs), and long short-term memory (LSTMs) NNs, the LSTM provided the best equalization performance on both simulated and experimental data. 
In another study, a bidirectional neural network was employed for receiver-side equalization of SOA-induced distortions in a coherent  $30\,\mathrm{Gbaud}$ 16-QAM back-to-back system \cite{Safarnejadian2025neural}, yielding a $4\,\mathrm{dB}$ Q-factor improvement over standard minimum mean-square error (MMSE) equalization without requiring prior knowledge of the SOA parameters.
Of particular interest are low-complexity neural network equalizers that can be implemented in real time on field-programmable gate arrays (FPGAs). For example, a two-layer perceptron with 33 and 14 neurons was demonstrated in a $50\,\mathrm{Gb/s}$ passive optical network (PON) \cite{kaneda2020fpga}, outperforming a maximum likelihood sequence estimator at the expense of increased hardware complexity. 
An FPGA-based MLP was also reported for a $100\,\mathrm{Gb/s}$ IM/DD PON back-to-back system using an SOA as a power booster \cite{roshanshomal2025low}, in which the authors employed a two-layer model (12 and 8 neurons) with a 32-symbol input window. 
Alternatively, a two-layer 1D CNN was applied in a $20\,\mathrm{GBaud}$ IM/DD PON \cite{ney2024achieving}, using longer input sequences (256 and 512 symbols), with 4 and 8 filters per layer, while jointly predicting 8 symbols.  
In a related (SOA-free) long-haul scenario (dual-polarization 16-QAM at $34\,\mathrm{Gbaud}$ over 14 spans of $70\,\mathrm{km}$), a biLSTM and a deep CNN were investigated \cite{freire2023implementing}. Both models used 81-symbol inputs with four channels (I/Q components from both polarizations), and the CNN comprised two convolutional layers with 70 and 2 filters, producing 61 symbol decisions simultaneously, while the biLSTM used 35 hidden units and a similar 2-filter convolutional decision layer. However, these FPGA implementations were evaluated only via offline simulations.
These examples illustrate that only simple, shallow MLPs and CNNs are suitable for real-time implementation, while recurrent architectures impose additional limitations on online use. Although MLPs and CNNs demonstrate lower performance when processing channels with memory, they can still be effectively applied for equalization in low-dispersion scenarios.

The advantages of SOAs -- including broadband amplification, cost-effectiveness, compactness, and integrability -- motivate researchers to apply them in long-haul, metro, and access networks. However, SOAs' nonlinear distortions caused by gain saturation and a high noise figure have limited their use. The former can now be effectively addressed with NN equalizers capable of learning impairment patterns. Moreover, it has been demonstrated that the simplest NNs can be implemented on an FPGA, enabling real-time equalization. In this paper, we perform a numerical simulation of the long-haul coherent communication system with a cascade of SOAs. We use a simplified multilayer perceptron to compensate for the combined fiber and SOA-induced nonlinear distortions. We keep the complexity of the NNs closer to that of those implemented on an FPGA to ensure real-time feasibility.

\section{System and components modelling}
\begin{figure}[t]
    \centering
    \includegraphics[width=0.99\linewidth]{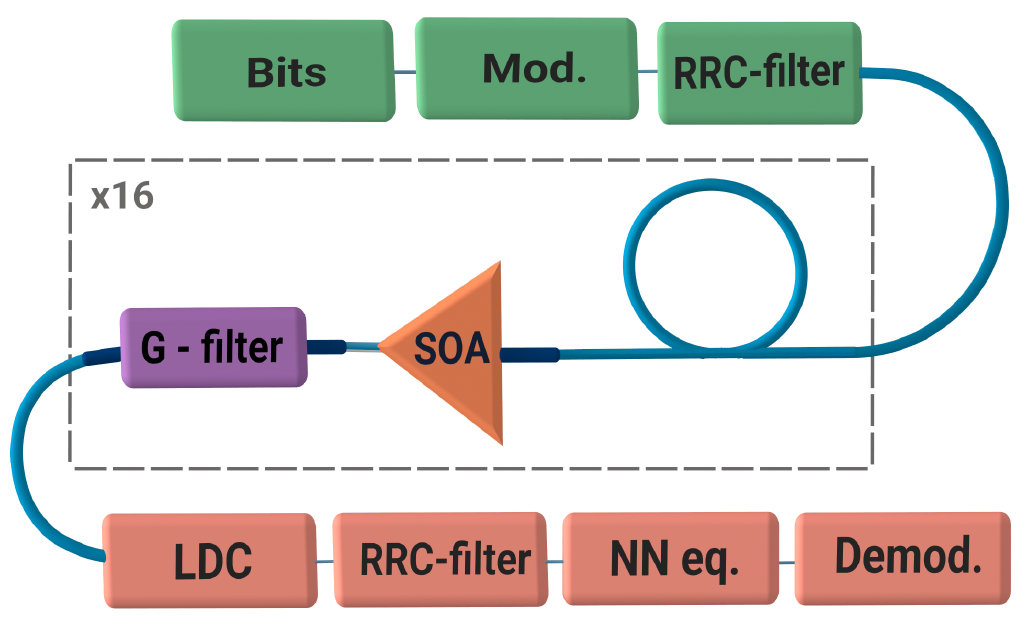}
    \caption{The scheme of the simulated data transmission system. At the transmitter: bit sequence generator, modulator, and RRC filter; in line: 16 spans each consisting of 73 km of standard fibre followed by an SOA and a Gaussian filter; at the receiver: linear dispersion compensation (LDC), matched RRC filter, NN equalizer, and demodulator.}
    \label{fig:sys}
\end{figure}

We consider single-polarization, single-channel transmission of a standard 16-QAM signal at $\mathrm{32\, Gbaud}$ over a multi-span fibre link, as depicted in Fig.~\ref{fig:sys}. At the transmitter, a bit sequence is mapped to 16-QAM symbols and shaped by a root-raised cosine (RRC) filter with roll-off 0.1. The waveform is then normalized to a prescribed launch power and injected into the transmission line. The link consists of 16 identical spans of standard single-mode fibre, each $\mathrm{73\,km}$ long, followed by an inline semiconductor optical amplifier (SOA) that compensates the span loss. Consequently, the signal power is restored after each span, while impairments due to fibre propagation and SOA operation (gain saturation and amplified spontaneous emission noise) accumulate along the cascade. After each SOA, a Gaussian optical filter is applied to suppress out-of-band amplified noise. At the receiver, we perform linear dispersion compensation (LDC), apply the matched RRC filter, downsample to $\mathrm{1\,Sa/sym}$, and finally apply a neural-network (NN) equalizer prior to demodulation.

Signal propagation in the optical fibre is governed by the nonlinear Schr\"odinger equation (NLSE) \cite{agrawal2000nonlinear}:
\begin{equation}\label{eq:NLSE}
    i q_z - \frac{\beta_2}{2}q_{tt} + \gamma |q|^2 q = -i \frac{\alpha_{\mathrm{f}}}{2} q,
\end{equation}
where $q(t,z)$ is the signal envelope, $t$ and $z$ are the temporal and spatial variables, $\beta_2$ is the group-velocity dispersion, $\gamma$ is the Kerr nonlinearity, and $\alpha_{\mathrm{f}}$ is the fibre attenuation. For standard single-mode fibre at $\lambda_0=1550\,\mathrm{nm}$ (C-band), we use $\beta_2=-21.67\,\mathrm{ps^2/km}$, $\gamma=1.27\times10^{-3}\,\mathrm{mW^{-1}km^{-1}}$, and $\alpha_{\mathrm{f}}=0.2\,\mathrm{dB/km}$. \eqref{eq:NLSE} is solved using the split-step Fourier method with 8 samples per symbol and 128 steps per span.

A comprehensive analysis of SOA-induced self-phase modulation and gain dynamics was presented in \cite{agrawal1989self}. For numerical simulations, we adopt a commonly used simplified (effectively lossless) dynamical SOA model, in which the gain is governed by a 1D rate equation and amplified spontaneous emission (ASE) noise is injected at the SOA input \cite{cassioli2000time}. More detailed approaches (e.g., traveling-wave and Connelly-type models) can be found in \cite{connelly2007semi, sobhanan2022semiconductor}. The SOA action on the complex envelope is described through the input/output power and phase relations \cite{agrawal1989self}:
\begin{equation}\label{eq:power}
    P_{\mathrm{out}}(t) = P_{\mathrm{in}}(t)\, e^{h(t)},
\end{equation}
\begin{equation}\label{eq:phas}
    \phi_{\mathrm{out}}(t) = \phi_{\mathrm{in}}(t) - \frac{1}{2}\alpha_{\mathrm{H}}\, h(t),
\end{equation}
\begin{equation}\label{eq:gain}
    \frac{\partial h(t)}{\partial t} =
    \frac{h_{0}-h(t)}{\tau_{c}}
    - \frac{P_{\mathrm{in}}(t)}{\tau_{c} P_{\mathrm{sat}}}\left(e^{h(t)}-1\right),
\end{equation}
where $P_{\mathrm{in}}(t)$ and $P_{\mathrm{out}}(t)$ are the instantaneous input and output powers, $\phi_{\mathrm{in}}(t)$ and $\phi_{\mathrm{out}}(t)$ are the corresponding phases, and $h(t)$ is the gain integrated over the SOA length. Here, $h_0$ is the small-signal integrated gain, $P_{\mathrm{sat}}$ is the saturation power, $\tau_c$ is the carrier lifetime, and $\alpha_{\mathrm{H}}$ is the linewidth enhancement factor. \eqref{eq:phas} shows that time-varying gain induces phase modulation and thus a gain-saturation-driven nonlinearity.

\begin{figure}[t]
    \centering
    \includegraphics[width=1.0\linewidth]{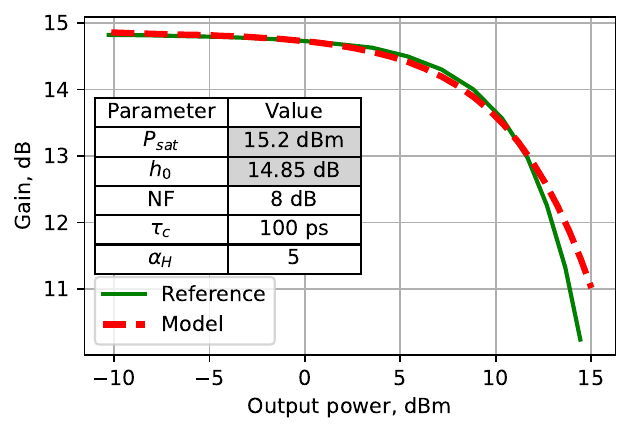}
    \caption{Comparison of the SOA model with the experimental data from Thorlabs' official webpage for SOA1013S \cite{thorlabs_SOA}.}
    \label{fig:SOA1013S_fit}
\end{figure}

We simulate the SOA using \eqref{eq:power}--\eqref{eq:gain}, solving \eqref{eq:gain} with a fourth-order Runge--Kutta method. ASE noise is added at the SOA input \cite{cassioli2000time}:
\begin{equation}
    q_{\mathrm{in}}^{\mathrm{SOA}}(t) = q(t) + \sigma_{\mathrm{ASE}}\, n_{\mathrm{ASE}}(t),
\end{equation}
where $q(t)$ is the signal before the SOA, $n_{\mathrm{ASE}}(t)$ is a complex random process normalized to unit power, and $\sigma_{\mathrm{ASE}}^2=h\nu n_{sp} B_{\mathrm{sim}}$. Here, $h\nu$ is the photon energy, $n_{sp}=\mathrm{NF}/2$ is the spontaneous emission factor with $\mathrm{NF}$ being the noise figure in linear units, and the simulation bandwidth is set by the temporal resolution $B_{\mathrm{sim}}=1/dt$. To mimic a realistic device, we tune the model parameters using a commercially available SOA1013S (Thorlabs) \cite{thorlabs_SOA}. The static CW gain curve provided by the vendor is compared with the model fit in Fig.~\ref{fig:SOA1013S_fit}. We adjust $h_0$ and $P_{\mathrm{sat}}$ to match the reference curve, while fixing $\mathrm{NF}=8\,\mathrm{dB}$ according to the device specification. The parameters $\tau_c$ and $\alpha_{\mathrm{H}}$ are not provided; therefore, we use typical values reported for SOAs. 

As a receiver-side equalizer, we employ a multilayer perceptron (MLP) to predict transmitted symbols using the sequences of received symbols. The MLP hyperparameters are selected via grid search. The NN input consists of a window of 33 consecutive complex-valued symbols; their I- and Q-components are concatenated into a real-valued vector before being fed to the network. The baseline model has one hidden layer with 128 neurons and the \texttt{tanh} activation function, and two outputs predicting the I and Q components of the central symbol; we refer to this architecture as \texttt{33-128}. In addition, we consider \texttt{17-64} and \texttt{33-256} variants. These architectures are intentionally kept close to those demonstrated in FPGA-oriented implementations reported in the literature \cite{kaneda2020fpga, roshanshomal2025low, ney2024achieving, freire2023implementing}, supporting the feasibility of real-time deployment. To train the NNs, we use $10^6$ symbols, split 80\% for training and 20\% for testing (when the BER is too low, we increase the test data to $10^6$ symbols), and the mean-squared-error loss function. The training routines are implemented in the TensorFlow 2.0 framework.

\section{Results}

\begin{figure}[b!]
    \centering
    \includegraphics[width=0.99\linewidth]{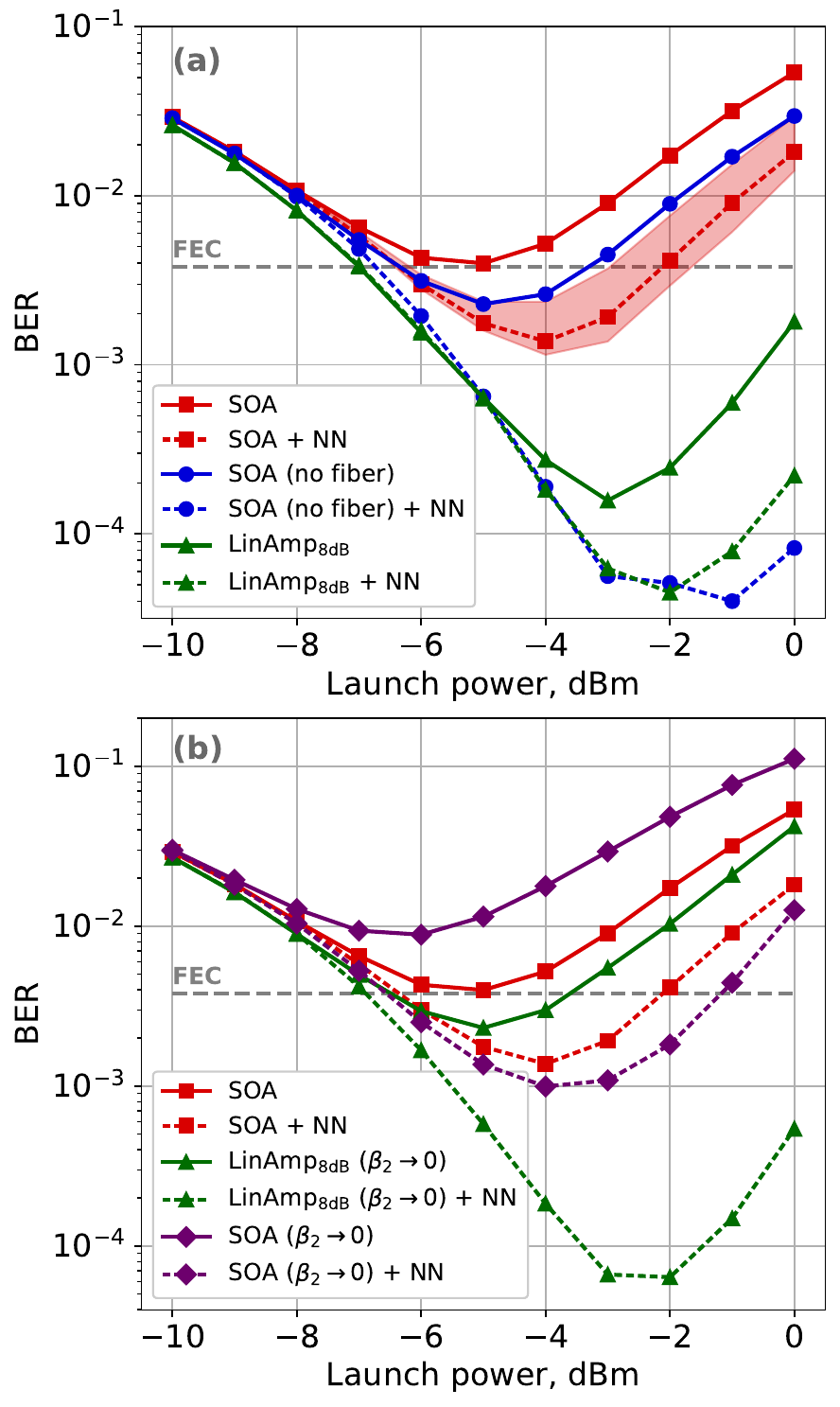}
    \caption{BER as a function of launch power for the system with SOA and $\mathrm{LinAmp_{8dB}}$ (linear amplifier with $\mathrm{NF=8\, dB}$). (a): The scenarios are SOA (red), SOA with no fiber (blue), $\mathrm{LinAmp_{8dB}}$ (green). (b): SOA (red, from (a) for reference), SOA with $\beta_2 \rightarrow 0$ (purple), and $\mathrm{LinAmp_{8dB}}$ with $\beta_2 \rightarrow  0$ (green). In all cases, solid lines correspond to non-equalized systems, dashed lines to equalized systems. The forward error correction (FEC) threshold is $3.8 \times 10^{-3}$ (7\% overhead).}
    \label{fig:SOA_1ch}
\end{figure}

In Fig.~\ref{fig:SOA_1ch}(a), we implement the data transmission system described above with inline SOAs and compare the performance with (1) a system in which the SOAs are replaced by linear amplifiers (with no any nonlinear dynamic) but the same $\mathrm{NF=8\,dB}$ (noted as $\mathrm{LinAmp_{8dB}}$), thereby excluding SOA nonlinearity from consideration. This allows us to assess how the NN compensates for deterministic SOA-induced effects, beyond the stochastic impairments caused by noise. The second scenario (2) is when the fiber spans between SOAs are replaced with variable optical attenuators (i.e., eliminating the fiber's dispersion and nonlinearity). We implement this regime to estimate the degree of distortion caused by a pure SOA cascade and to examine how the NN handles it. When we do not apply the NN equalizer, the SOAs' nonlinearity is the main contributor to the signal distortion.
However, when the NN, \texttt{33-128} model, is incorporated into the DSP chain, we observe that distortions in the system with removed fiber spans, the scenario (2), could be effectively compensated, 57x reduction in BER. At the same time, a limited performance gain is achieved in the complete system (including fiber spans), 3x lower BER. Even with different NN models from our pool, we receive almost the same performance (red shaded area in Fig.~\ref{fig:SOA_1ch}(a)).
Also, a 3.5x decrease in BER is achieved with the NN equalizer in the scenario where SOAs are replaced with $\mathrm{LinAmp_{8dB}}$. Therefore, it is clear that the joint action of the SOAs' nonlinearity and fiber effects significantly reduces the performance gain from the NN equalizer, whereas the pure SOAs' distortions can be effectively compensated.

\begin{figure}[t]
    \centering
    \includegraphics[width=0.99\linewidth]{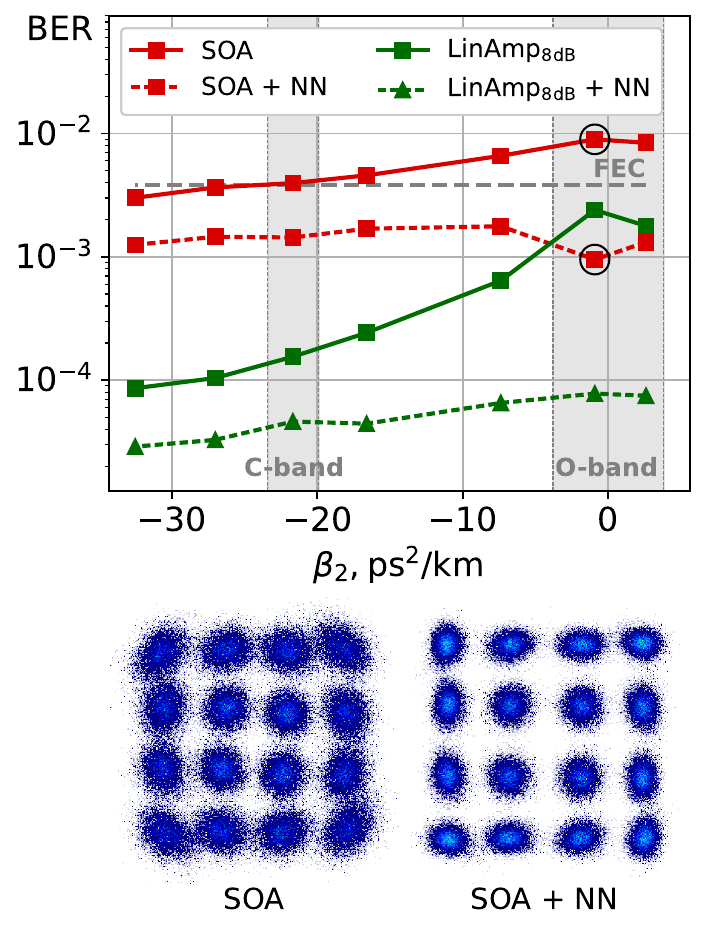}
    \caption{The dependence of BER on $\beta_2$ for system with SOA and $\mathrm{LinAmp_{8dB}}$ when NN is applied or not. Two constellations correspond to the wavelength at which NN equalization provides the maximum BER reduction (circled on the graph).}
    \label{fig:SOA_over_bands}
\end{figure}

The dispersion of the optical fiber introduces channel memory, complicating the equalization of signals with NNs. In Fig.~\ref{fig:SOA_1ch}(b), we compare our communication system having inline SOAs with one where the optical fiber dispersion is set to be small, $\beta_2 = \mathrm{-1\, ps^2/km}$ (depicted $\beta_2 \rightarrow 0$ in Fig.~\ref{fig:SOA_1ch}(b)). The last scenario can be applied to the O-band data transmission. Although systems with small dispersion suffer from increased nonlinear distortion, the application of NNs benefits by reducing channel memory. In this case, our NN equalizer provides a 9x reduction in BER, compared to 3x for the system with standard C-band dispersion ($\beta_2 = \mathrm{-21.7\, ps^2/km}$).
The same holds for the SOAs replaced by $\mathrm{LinAmp_{8dB}}$ at the same small dispersion regime, where the NN equalizer reduces BER by up to 36x. These findings reveal that the nonlinear distortions caused by the joint action of an SOA and fiber nonlinearity can be effectively compensated by a simple NN equalizer when isolated from dispersion.

The dependence of the gain in performance for systems with SOAs and $\mathrm{LinAmp_{8dB}}$ on a wide range of $\beta_2$ after applying the NN equalizer is depicted in Fig.~\ref{fig:SOA_over_bands}. It is clearly demonstrated that the NN is able to provide better equalization when $|\beta_2|$ is small. For the system with $\mathrm{LinAmp_{8dB}}$, this gain is even higher, demonstrating that NN compensates not only SOAs' but fiber's nonlinearity as well. 

\section{Conclusion}
In this letter, we investigate a single-polarization single-channel coherent long-haul data transmission system with a cascade of 16 SOAs as inline amplifiers.
We demonstrate that the nonlinear distortions of a signal caused by a cascade of SOAs can be effectively compensated by a low-complexity NN, enabling real-time implementation. Moreover, the proposed equalizer operates effectively when SOA's distractive effects are mixed with fiber nonlinearity.
At the same time, the dispersion of the optical fiber introduces channel memory, significantly reducing equalization performance.
However, as $|\beta_2|$ tends to zero (i.e., in the O-band), the effectiveness of equalization increases by an order of magnitude.

\section{Acknowledgment} This work was supported by UK EPSRC grant TRANSNET (EP/R035342/1) and Advanced Optical Frequency Comb Technologies and Applications (EP/W002868/1). 

% Bibliography
\bibliography{bibliography}

\end{document}